\documentclass[twocolumn]{aastex62}
\usepackage{amsmath,amssymb,graphicx}

\newcommand{\cc}{\text{cm}^{-3}}
\newcommand{\cmmm}{\text{cm}^{-3}}

\newcommand{\msunyr}{\text{M}_\odot\,\text{yr}^{-1}}
\newcommand{\K}{\rm{K}}
\newcommand{\rB}{R_{\rm B}}
\newcommand{\nB}{n_{\rm B}}
\newcommand{\nstar}{n_\star}

\newcommand{\HH}{\text{H}_2}          
\newcommand{\HM}{{\rm H}^{-}}     
\newcommand{\HP}{{\rm H}^{+}}     
\newcommand{\e}{{\rm e}^{-}}     

\newcommand{\mj}{M_{\rm J}}

\newcommand{\msun}{\text{M}_\odot}

\newcommand{\tff}{t_{\rm ff}}

\newcommand{\Tvir}{T_{\rm vir}}

\newcommand{\cs}{c_{\rm s}}
\newcommand{\css}{c_{\rm s}^2 }

\newcommand{\vrad}{v_{\rm rad}}

\newcommand{\mstardot}{\dot{M}_\star}
\newcommand{\mstar}{M_\star}
\newcommand{\rstar}{R_\star}
\newcommand{\tstar}{T_\star}

\newcommand{\mh}{m_{\rm H}}
\newcommand{\kb}{k_{\rm B}}
\newcommand{\Racc}{R_{\rm acc}}
\newcommand{\mf}{M_{\rm F}}
\newcommand{\nf}{n_{\rm F}}
\newcommand{\rf}{R_{\rm F}}
\newcommand{\nth}{n_{\rm th}}
\newcommand{\mdot}{\dot{M}}

\newcommand{\Tion}{T_{\rm ion}}
\newcommand{\tacc}{t_{\rm acc}}
\newcommand{\tkh}{t_{\rm KH}}

\newcommand{\Lhf}{\Lambda^{(h)}_{\rm thin}}
\newcommand{\Llf}{\Lambda^{(l)}_{\rm thin}}
\newcommand{\khf}{k^{(h)}}
\newcommand{\klf}{k^{(l)}}
\newcommand{\klfffP}{\kappa^{(l)}_{\rm ff, P}}
\newcommand{\khfffP}{\kappa^{(h)}_{\rm ff, P}}
\newcommand{\khfbfP}{\kappa^{(h)}_{\rm bf, P}}
\newcommand{\klfffR}{\kappa^{(l)}_{\rm ff, R}}
\newcommand{\khfRay}{\kappa^{(h)}_{\rm Ray}}
\newcommand{\khfR}{\kappa^{(h)}_{\rm R}}
\newcommand{\klfR}{\kappa^{(l)}_{\rm R}}
\newcommand{\khfP}{\kappa^{(h)}_{\rm P}}
\newcommand{\klfP}{\kappa^{(l)}_{\rm P}}

\accepted{to ApJ}

\shorttitle{Opacity limit for supermassive protostars}
\shortauthors{Becerra et al.}

\begin{document}

\title{Opacity limit for supermassive protostars}

\correspondingauthor{Fernando Becerra}
\email{fbecerra@cfa.harvard.edu}

\author{Fernando Becerra}
\affil{Harvard-Smithsonian Center for Astrophysics, 60 Garden Street, Cambridge, MA 02138, USA}

\author{Federico Marinacci}
\affiliation{Kavli Institute for Astrophysics and Space Research, Massachusetts Institute of Technology, Cambridge, MA 02139, USA}

\author{Kohei Inayoshi}
\affiliation{Department of Astronomy, Columbia University, 550 W. 120th Street, New York, NY 10027, USA}

\author{Volker Bromm}
\affiliation{Department of Astronomy, The University of Texas at Austin, TX 78712, USA}

\author{Lars E. Hernquist}
\affiliation{Harvard-Smithsonian Center for Astrophysics, 60 Garden Street, Cambridge, MA 02138, USA}

\begin{abstract}

We present a model for the evolution of supermassive protostars from their formation at $\mstar \simeq 0.1\,\msun$ until their growth to $\mstar \simeq 10^5\,\msun$. To calculate the initial properties of the object in the optically thick regime we follow two approaches: based on idealized thermodynamic considerations, and on a more detailed one-zone model. Both methods derive a similar value of $\nf \simeq 2 \times 10^{17} \,\cmmm$ for the density of the object when opacity becomes important, i.e. the opacity limit. The subsequent evolution of the growing protostar is determined by the accretion of gas onto the object and can be described by a mass-radius relation of the form $\rstar \propto \mstar^{1/3}$ during the early stages, and of the form $\rstar \propto \mstar^{1/2}$ when internal luminosity becomes important. For the case of a supermassive protostar, this implies that the radius of the star grows from $\rstar \simeq 0.65 \,{\rm AU}$ to $\rstar \simeq 250 \,{\rm AU}$ during its evolution. Finally, we use this model to construct a sub-grid recipe for accreting sink particles in numerical simulations. A prime ingredient thereof is a physically motivated prescription for the accretion radius and the effective temperature of
the growing protostar embedded inside it. From the latter, we can conclude that photo-ionization feedback can be neglected until very late in the assembly process of the supermassive object.

\end{abstract}


\section{Introduction}
\label{sec:intro}

Recent observations at redshifts $z \gtrsim 6$ suggest that quasars were already powered by supermassive black holes (SMBHs) with masses $\gtrsim 10^9\,\msun$ when the Universe was less than one billion years old \citep{Fan_2003, Fan_2006, Mortlock_2011, Wu_2015}. These SMBHs most likely grew from smaller seed BHs that formed earlier, but the origin of these seeds remains unclear \citep{Haiman_2006, Haiman_2009, Bromm_Yoshida_2011, Greene_2012, Volonteri_2012, Volonteri_Bellovary_2012}. Furthermore, feedback and self-regulation of the seeds make the study of their formation and growth even more complex \citep{Milosavljevic_2009}. The two most promising theories concerning the formation of seed BHs at high redshift are the remnants of massive Population~III stars \citep{Madau_2001, Li_2007, Johnson_2012}, and the direct collapse of primordial gas in haloes with virial temperatures $T_{\rm vir}\gtrsim 10^4\,$K, the so-called atomic cooling haloes \citep{Bromm_2003, Begelman_2006, Spaans_2006}. 

\begin{figure*}
\begin{center}
\includegraphics[scale=0.25]{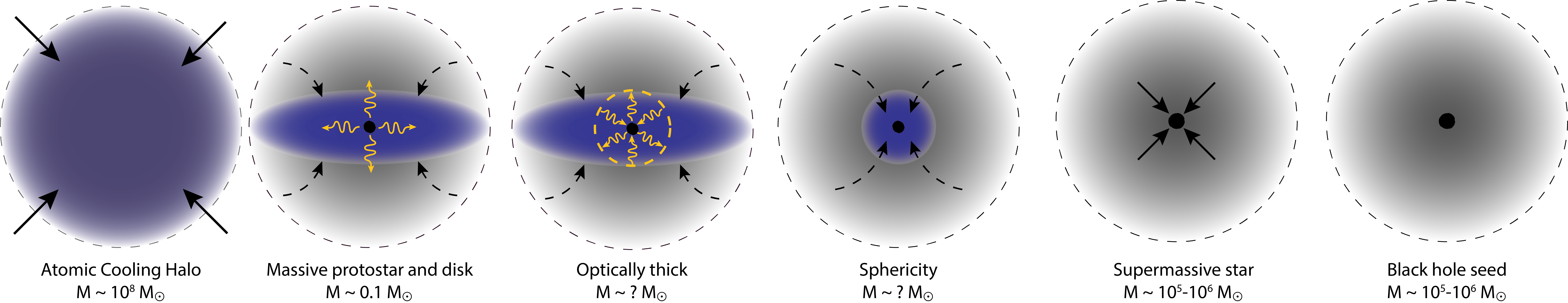}
\caption{Overview of the formation of a supermassive black hole seed. An atomic cooling halo of virial (total) mass $M \simeq 10^8 \msun$ and exposed to strong Lyman-Werner background radiation collapses. The gas reaches the optically thick regime first on small scales, such that a central protostar of initial mass $M \simeq 0.1\,\msun$ and accretion rate $\mdot \simeq 1\,\msunyr$ is formed, surrounded by a disk-like structure. Photons coming from the protostar due to accretion are radiated away, until the gas becomes optically thick to $\HM$ radiation at intermediate scales. Eventually, the central object eats up the entire disk and tends toward sphericity, although the mass of the object at each of these later stages still remains to be determined. The massive protostar keeps accreting the surrounding gas and becomes a supermassive star of $M \simeq 10^5-10^6\,\msun$ after $\simeq 10^5-10^6\,{\rm yr}$. Finally, it collapses into a massive black hole seed due to relativistic instabilities.}
\label{fig:sketch}
\end{center}
\end{figure*}

In the direct collapse scenario, high temperatures are reached in haloes where cooling by molecular hydrogen and metal lines to below $\simeq 10^4\,$K has been suppressed, which implies that the only coolant acting on the gas is atomic hydrogen \citep{Omukai_2001, Oh_2002}. In the case of molecular hydrogen, which naturally forms at the center of the halo, its photo-dissociation can be achieved by an external soft ultraviolet (UV) background in the Lyman-Werner (LW) bands. Previous studies have found that this leads to a nearly isothermal collapse at $\Tvir \simeq 10^4\,{\rm K}$ due to initially Lyman-$\alpha$ cooling, and subsequently $\HM$ bound-free and free-free emission, when higher densities are reached \citep{Regan_2009, Latif_2013a, Inayoshi_2014, Becerra_2015, Chon_2016}. High-resolution simulations have shown that, as the gas collapses and reaches densities of $\simeq 10^{17} \cmmm$, it becomes optically thick to $\HM$ radiation, and a massive protostar with accretion rate $\simeq 1\,\msunyr$ forms at the center of the halo \citep{Inayoshi_2014, VanBorm_2014, Becerra_2015, Latif_2016}. Due to this high accretion rate, the central object can easily become a supermassive star of $\simeq 10^5 - 10^6\,\msun$ within a million years \citep{Regan_2009, Latif_2013a}, which later might collapse into a SMBH due to relativistic instabilities \citep[][see also Figure \ref{fig:sketch}]{Baumgarte_1999, Umeda_2016, Woods_2017}.

In this work, we study the physics of the central object when it approaches the optically thick regime. In particular, we investigate the properties of the emerging protostar when the optical depth due to $\HM$ emission becomes unity, thus extending the classical theory of opacity-limited fragmentation developed for present-day star formation \citep{Rees_76, Low_76}. Previous studies have explored this scenario using detailed one-zone models \citep[e.g.][]{Omukai_2001}. Here, we present an alternative approach based on both simplified dimensional arguments and a fitting formula for the cooling and heating processes within the non-equilibrium chemistry of H and $\HM$ ions. In addition, we develop an idealized model for the subsequent evolution of the accreting protostar, until the formation of a supermassive object. Based on this modeling of the growing protostar, we deduce parameters for a physically-motivated sink particle algorithm, to be used as a sub-grid recipe in large-scale, hydrodynamic simulations of the formation of SMBH seeds in a fully cosmological context. Such simulations are needed to derive detailed diagnostics for the SMBH assembly process at high redshifts, to be probed with next-generation observational facilities \citep{Pacucci_2015}, such as the {\it James Webb Space Telescope (JWST)}, the ATHENA X-ray mission, and the Laser Interferometer Space Antenna (LISA) gravitational-wave observatory.

\section{Physics of the opacity limit}
\label{sec:physics}

\subsection{Classical picture}
\label{subsec:classical}

In the theory of star formation, it has been a long-standing quest to understand the limits to fragmentation in a given cloud setting. An influential idea was that fragmentation proceeds hierarchically in a collapsing cloud, as the Jeans mass decreases with increasing density as long as the cloud can collapse almost isothermally \citep{Hoyle_1953}. The minimum fragment mass is then set by the scale when opacity prevents the release of the gravitational energy via radiation, such that the Jeans mass would increase again upon further compression \citep{Rees_76, Low_76}. Fragmentation can be seeded in a number of ways, including from non-spherical perturbations of the Larson-Penston solution \citep{Hanawa_2000, Lai_2000}. We here follow a similar reasoning, applied to the peculiar conditions of isothermally collapsing primordial gas in atomic cooling haloes. Our goal is to robustly derive the characteristic density $\nf$, mass $\mf$ and radius $\rf$ of the emerging protostar, when the gas first becomes optically thick (see third panel of Figure \ref{fig:sketch}). These values will mark the initial stage in the build-up process of the supermassive object.

We start by considering the simple relation between these three quantities:
\begin{equation}
\mf = {4\pi \over 3} {\mh \over X} \nf \rf^3
\mbox{\ ,}
\label{eq:density}
\end{equation}
\noindent where we have used $\rho = \mh n / X$, with $X=0.76$ being the primordial hydrogen mass fraction, to translate total mass density to hydrogen number density.

We furthermore assume that the optically thick cloud is gravitationally bound, such that the characteristic mass is of order the Jeans mass $\mj$,
\begin{equation}
\mf \simeq \mj = \left( \pi \kb \over G \mh \mu \right)^{3/2} T^{3/2} \nf^{-1/2}
\mbox{\ ,}
\label{eq:jeans_mass}
\end{equation}
where $\mu \simeq 1.22$ is the mean molecular weight for a fully-neutral primordial mixture 
of atomic hydrogen and helium.

Finally, we need to account for energy equilibrium. 
The energy that is to be radiated away originates in the gravitational collapse of the cloud. 
In this case, the gravitational energy is emitted in a collapse timescale $t_{\rm col}$, as long as the gas remains optically thin to its cooling radiation. 
Right before the gas cloud becomes opaque, the energy is radiated from the surface as a fraction $f_{\rm BB}$ of the black-body radiation.
Hence, we can equate the gravitational compressional heating rate with the radiation cooling rate
\begin{equation}
{4\pi \over 3} \rf^3 \cdot {\nf \kb T \over t_{\rm col}} = f_{\rm BB} 4 \pi \rf^2 \sigma_{\rm SB} T^4
\mbox{\ ,}
\label{eq:energy}
\end{equation}
\noindent where $\sigma_{\rm SB}$ is the Stefan-Boltzmann constant. In a one-zone model, where the thermal evolution of the central core of a collapsing cloud is calculated,
the collapse timescale is commonly assumed to be $t_{\rm col}=\sqrt{3\pi/32G \rho}$, which 
is the time for the density of an initially static cloud to reach infinity.
On the other hand, the dynamical timescale in the free-fall collapse is $t_{\rm col}=1/\sqrt{24G \rho}$, 
which is shorter by a factor of $3\pi/2\simeq4.7$.
In reality, the collapse timescale can be between these values.
Here, we set the collapse timescale to $t_{\rm col}=f_{\rm col} \sqrt{3\pi/32G \rho}$ in order to consider this uncertainty, where $0.2\la f_{\rm col} \la 1$.

We then proceed to solve the system of equations (\ref{eq:density}), (\ref{eq:jeans_mass}), and (\ref{eq:energy}), and obtain analytic expressions of the characteristic density, radius and mass

\begin{equation}
n_{\rm F} \simeq 4.6 \times 10^{16}~f
\left(\frac{T}{3000~\K}\right)^{5/2}~\cc,
\label{eq:nF}
\end{equation}
\begin{equation}
R_{\rm F} \simeq 0.33~f^{-1}
\left(\frac{T}{3000~\K}\right)^{-3/4}~{\rm AU}
\label{eq:RF}
\end{equation}
\begin{equation}
\mf \simeq 0.045~f^{-1/2}
\left(\frac{T}{3000~\K}\right)^{1/4}~\msun
\mbox{\ ,}
\label{eq:MF}
\end{equation}
where we have normalized $T$ to the typical value of isothermally collapsing gas in the high density regime in an atomic cooling halo and we have used $f = f_{\rm col} f_{\rm BB} \lesssim 1$. We note that this argument for the universal line in the density-temperature plane (Equation \ref{eq:nF}), on which the gas cloud becomes optically thick to any continuum opacities (gas and dust grains) has already been discussed by \citet{Omukai_2005}, and we here reproduce the key result in a simplified way to highlight the basic physics involved.

\begin{figure}
\begin{center}
\includegraphics[scale=0.32]{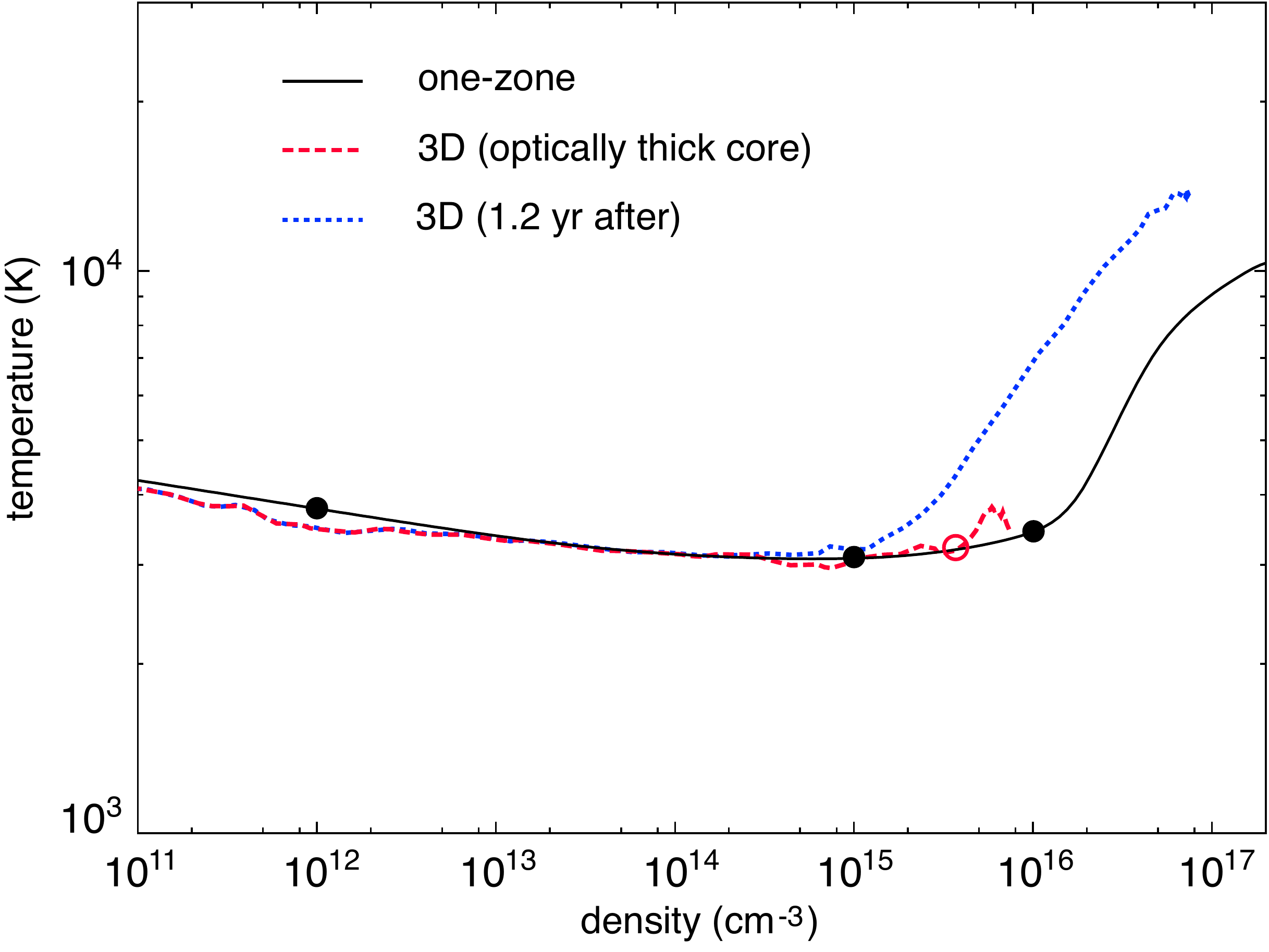}
\caption{Density-temperature diagram of a collapsing cloud in an atomic cooling halo.
Solid curve (black) presents the thermal history obtained from a one-zone calculation 
including H$^-$ continuum cooling and opacity effects (see Section \ref{subsec:onezone}).
Dashed curves (red and blue, respectively) show the snapshots of 
a three-dimensional (3D) simulation by \cite{Inayoshi_2014}
when (1) a hydrostatic protostar forms due to the opacity limit (long, red),
and (2) $1.2$ yr after protostellar formation (short, blue).
The open circle marks the density above which the gas becomes opaque in the 3D simulation.
Filled circles mark the three epochs at which we show the optical depth due to absorption and scattering
in Figure \ref{fig:tau}.
The density at the opacity limit in the 3D simulation is lower than that in the one-zone calculation
because the collapse timescale in the 3D simulation is shorter than what is assumed in the one-zone model. 
}
\label{fig:nt}
\end{center}
\end{figure}

Figure \ref{fig:nt} shows the density-temperature diagram of a collapsing cloud in an atomic cooling halo 
when a protostar, composed of a hydrostatic and adiabatic core, forms at the center of the cloud (red dashed curve).
The data is taken from a three-dimensional (3D) hydrodynamical simulation by \cite{Inayoshi_2014},
including all relevant cooling and chemical reaction networks.
In this case, the gas becomes opaque at $n\simeq 5\times 10^{15}~\cc$ (open circle),
which agrees with $n_{\rm F}=9.2\times 10^{15}~\cc$ for $f=0.2$ within a factor of two 
(see Equation \ref{eq:nF}).
Moreover, the radius and mass of the opaque core estimated from Equations (\ref{eq:RF}) and (\ref{eq:MF}) 
are $R_F=1.7$ AU and $M_F=0.1~\msun$ for $f=0.2$, respectively.
Both values also reasonably agree with the 3D simulation results, where $R_F\simeq 1$ AU and $M_\star \simeq 0.2~\msun$ right after
protostellar formation.

\begin{figure}
\begin{center}
\includegraphics[scale=0.32]{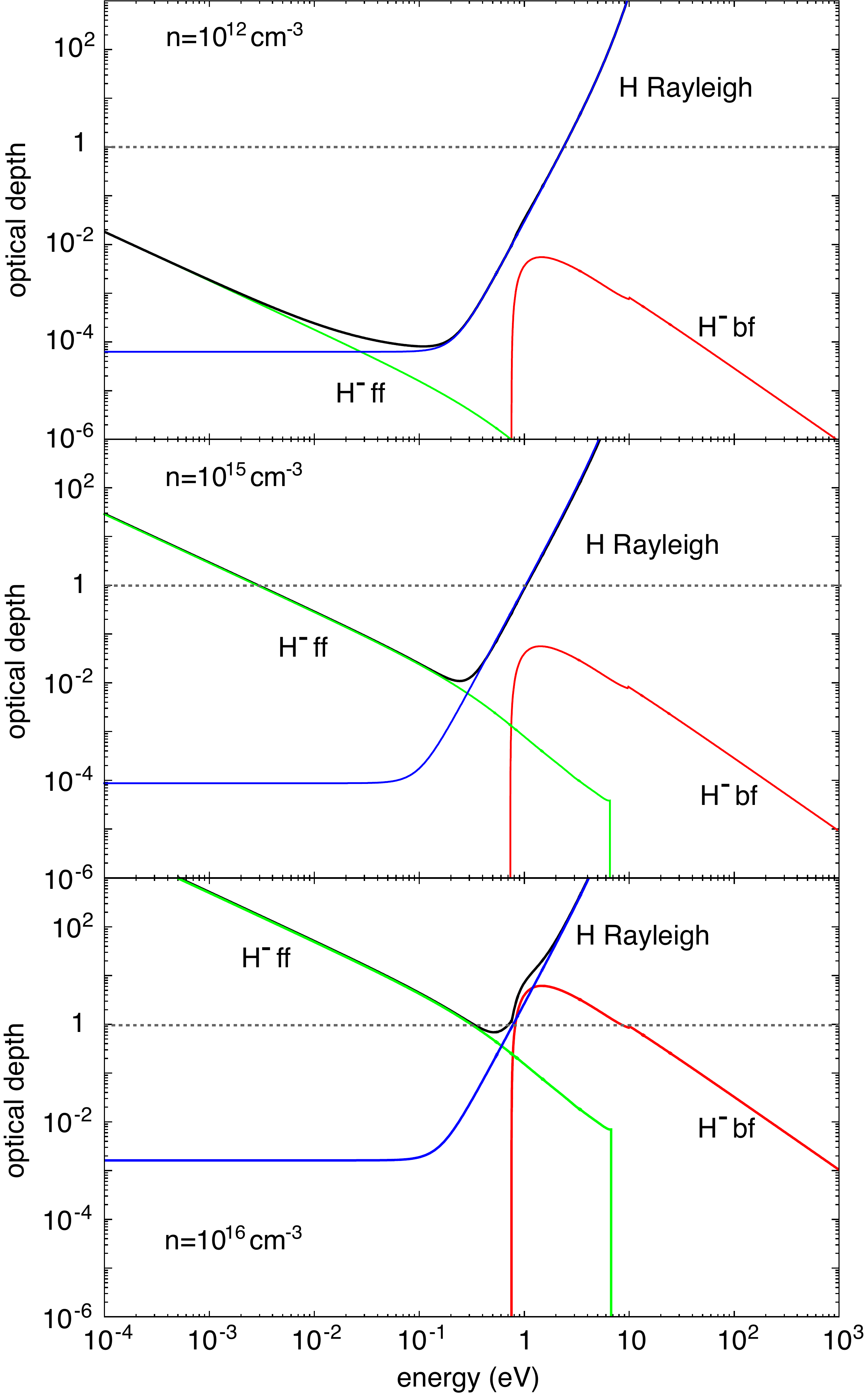}
\caption{Frequency-dependent optical depth due to H$^-$ bound-free (red), H$^-$ free-free (green) and H Rayleigh scattering (blue)
for three different densities with $n=10^{12}~\cc$ (top), $10^{15}~\cc$ (middle) and $10^{16}~\cc$ (bottom).
Horizontal dashed line shows the line on which $\tau_\nu (= n \sigma_\nu \lambda_{\rm J})=1$.}
\label{fig:tau}
\end{center}
\end{figure}

\subsection{Detailed modeling: One-zone model}
\label{subsec:onezone}

Next, we gain further insight by considering the detailed physics of $\HM$ opacity, 
based on the actual microphysical cross section for this process. 
To this extent, we re-derive the opacity limit by using a one-zone model for the collapse of gas into an atomic cooling halo. 

Following previous works \citep{Omukai_2001,Inayoshi_2014}, we implement the cooling function due to H$^-$ free-bound (fb)
and free-free (ff) emission
\begin{equation}
{\rm H}+{\rm e}^- \rightarrow {\rm H}^-+\gamma,
\end{equation}
\begin{equation}
{\rm H}+{\rm e}^- \rightarrow {\rm H}+{\rm e}^- + \gamma, 
\end{equation}
and consider three opacity sources associated with H$^-$ bound-free, free-free transition 
and H Rayleigh scattering 
\begin{equation}
{\rm H}^-+\gamma \rightarrow {\rm H}+{\rm e}^-,
\end{equation}
\begin{equation}
{\rm H}+{\rm e}^- +\gamma \rightarrow {\rm H}+{\rm e}^-,
\end{equation}
\begin{equation}
{\rm H}+\gamma \rightarrow {\rm H} + \gamma'.
\end{equation}
These processes are treated in a self-consistent way with chemical reaction networks.
An updated set of chemical reaction rate coefficients and cross sections is summarized 
in \cite{Inayoshi_2014,Inayoshi_2016}.

In what follows, we briefly describe the method introduced by \cite{Inayoshi_2014} to calculate the cooling function 
both in the optically thin and thick regime.
We summarize specific functional forms for the cooling rates and opacities in the Appendix.
In the optically thin limit, the $\HM$ cooling rate is estimated by integrating 
emissivities over frequency as
\begin{align}
\Lambda_{\rm thin} \equiv 4\pi \int (\eta_{\nu}^{\rm fb}+\eta_{\nu}^{\rm ff}) d \nu.
\label{eq:hmthin1}
\end{align}
We here divide the frequency range into two: 
$D_l=[0, 0.75]$ eV and $D_h=[0.75, 13.6]$ eV, 
called ``low" and ``high" frequency, respectively. That is, 
\begin{align}
\Lambda_{\rm thin} &=\int _{D_l}+\int _{D_h}
4\pi (\eta_{\nu}^{\rm fb}+\eta_{\nu}^{\rm ff}) d \nu, \nonumber \\
&\equiv \Lambda_{\rm thin}^{(l)}+\Lambda_{\rm thin}^{(h)}.
\label{eq:hmthin2}
\end{align}
This distinction between the two ranges is required to calculate the cooling rate 
in the optically thick case.
Figure \ref{fig:tau} shows the optical depth at the core of the collapsing cloud 
due to H$^-$ bound-free/free-free transition and H Rayleigh scattering 
for three different densities.
For the lowest density ($n=10^{12}~\cc$, top panel), the gas is optically thin ($\tau_\nu = n \sigma_\nu \lambda_{\rm J}<1$) to all the continuum opacities at frequencies $\la 2$ eV.
As the density increases to $n=10^{15}~\cc$ (middle panel), the optical depth at higher frequencies ($\ga 1$ eV) exceeds unity, 
but the H$^-$ free-free emission still works as radiation cooling.
For the highest density  ($n=10^{16}~\cc$, bottom panel), the gas core becomes completely opaque to all the continuum, and hence enters the opacity limit.

In the optically thick limit, the cooling function is approximated as 
\begin{equation}
\Lambda_{\rm thick} \simeq  \int _{D_l}+\int _{D_h}
\frac{-4\pi}{3(\kappa_\nu^{\rm a} +\kappa_\nu^{\rm s})}
\frac{\partial ^2 B_\nu (T)}{\partial z^2}d \nu,
\label{eq:cooling_1}
\end{equation}
where $\kappa_\nu^{\rm a(s)}$ is the absorption (scattering) coefficients,
$B_{\nu}(T)$ is the Planck function and $z$ is the coordinate along the temperature gradient.
Here, we approximate Equation (\ref{eq:cooling_1}) as 
\begin{equation}
\Lambda_{\rm thick} \simeq \sum_{i=l,h}
\frac{4\pi \int _{D_i}B_\nu(T)d \nu}{3\kappa_R^{(i)}\ell ^2}
=\sum_{i=l,h}
\frac{4\pi \int _{D_i}\eta_\nu d \nu}{3\kappa_R^{(i)}\kappa_P^{(i)}\ell ^2},
\label{eq:cooling_2}
\end{equation}
where the partial derivative $\partial/\partial z$ is replaced with a characteristic length $\ell$,
and $\kappa_R^{l(h)}$ and $\kappa_P^{l(h)}$ are Rosseland and Planck mean opacity
in the low and high frequency regime, respectively.
Note that, in this limit, the emissivity is expressed as $\eta_\nu=\kappa_\nu^{\rm a}B_\nu(T)$
because the source function is given by $B_\nu(T)$.

\begin{figure}
\begin{center}
\includegraphics[scale=0.32]{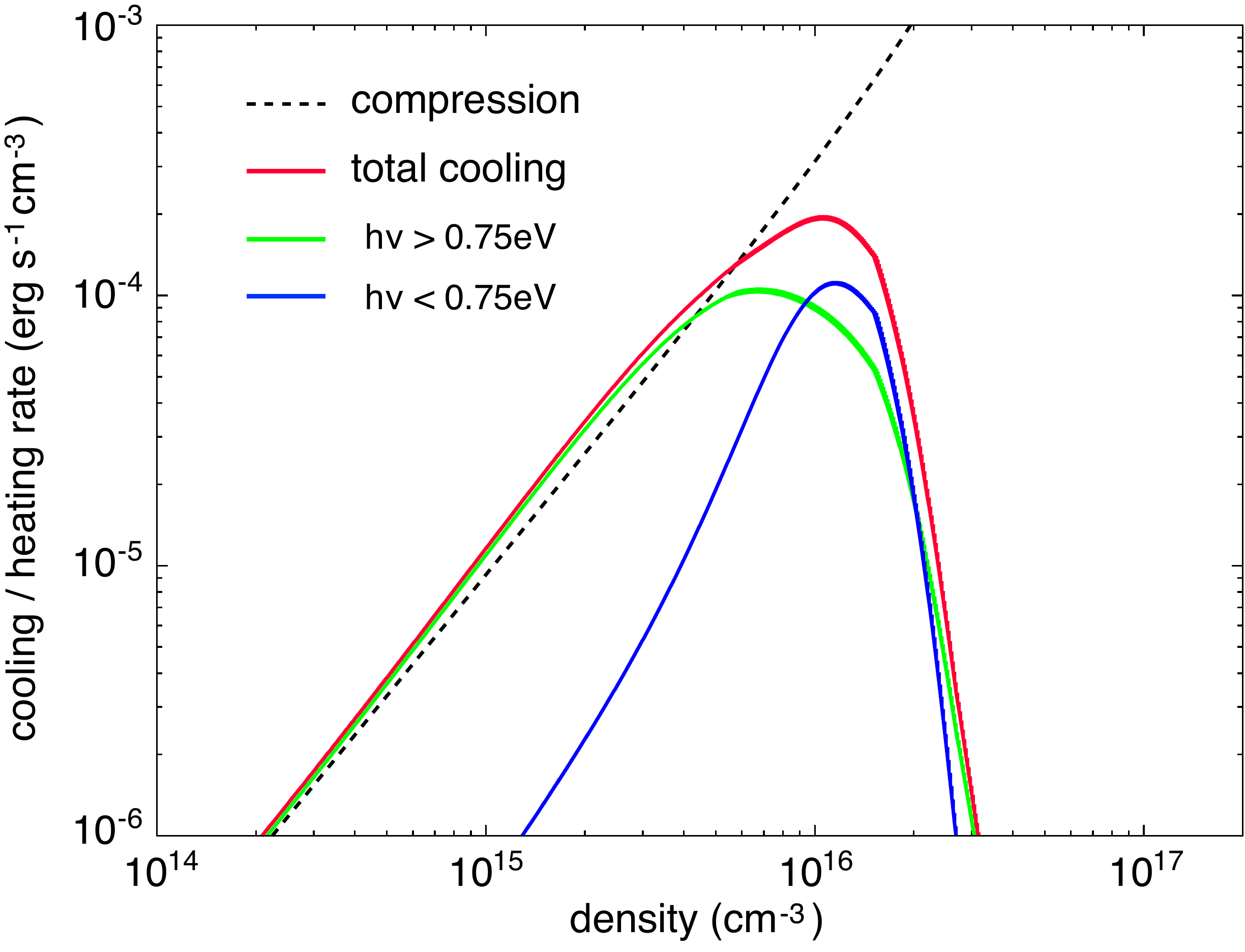}
\caption{Radiative cooling rates (solid) and heating rate due to gravitational compression (dashed)
in a collapsing cloud.
Cooling rates represented are the total cooling rate ($\Lambda_{\rm tot}$, red solid),
the rate due to higher-frequency photons with $h\nu>0.75$ eV ($\Lambda^{(h)}$, green solid) 
and with $h\nu<0.75$ eV ($\Lambda^{(l)}$, blue solid).
}
\label{fig:n_rate}
\end{center}
\end{figure}

Finally, in order to connect both the optically thin and thick regime, 
we adopt the following functional form
\begin{align}
\Lambda_{\rm tot} &\simeq \sum_{i=l,h}
\frac{4\pi \int _{D_i}\eta_\nu d \nu}{1+3\kappa_R^{(i)}\kappa_P^{(i)}\ell ^2},\nonumber\\
&=\sum_{i=l,h}
\frac{\Lambda _{\rm thin}^{(i)}}{1+3\kappa_R^{(i)}\kappa_P^{(i)}\ell ^2}
\equiv \Lambda^{(l)}+\Lambda^{(h)},
\label{eq:cooling_3}
\end{align}
where the first (second) term in the right-hand-side in Equation (\ref{eq:cooling_3})
mainly corresponds to H$^-$ free-bound (free-free) emission.
Figure \ref{fig:n_rate} shows the evolution of the cooling rates
for a collapsing cloud in our one-zone calculation (see solid curve in Figure \ref{fig:nt}).
Each solid curve presents the total cooling rate (red), the rate of $\Lambda^{(h)}$ (green) 
and $\Lambda^{(l)}$ (blue).
The H$^-$ free-bound cooling saturates and decreases at $n>6\times 10^{15}~\cc$ 
because of H Rayleigh scattering and H$^-$ bound-free absorption.
At $9\times 10^{15}~\cc \la n \la 2\times 10^{16}~\cc$, the H$^-$ free-free emission
acts as the main cooling process instead of H$^-$ free-bound. 
Since the compressional heating, given by $\Gamma_{\rm comp}=nk_{\rm B}T/t_{\rm ff}$ 
(dashed curve in Figure \ref{fig:n_rate}), dominates the total cooling rate during this transition,
the temperature begins to increase gradually.
Eventually, the gas becomes completely opaque at $n>2\times 10^{16}~\cc$, where $T\propto n^{2/3}$.
Note that this density, here derived by considering the detailed microphysics involved, is 
very similar to the estimate in Section~\ref{subsec:classical}. 
We can thus robustly characterize the conditions at the onset of supermassive protostar formation.

\subsection{Protostellar evolution}

After the collapse and formation of the optically thick object, its mass grows through accretion of the surrounding gas and new sources of energy start becoming important (see fourth panel of Figure \ref{fig:sketch}). In that case, its evolution will not be determined by the energy radiated away from the collapse, but by
the interplay between internal and accretion radiation from the protostar.
During the first stage of protostellar evolution, the energy powering 
the object will dominantly come from accretion rather than self-gravitating collapse.
At some point toward the later evolution of the system, the accretion timescale, 
$t_{\rm acc}=\mstar/ \mstardot$, becomes larger than the Kelvin-Helmholtz (KH) timescale, 
$t_{\rm KH} = G \mstar^2 / \rstar L_\star$, and hence the protostellar model 
needs to be augmented by internal contributions \citep[e.g.,][]{Omukai_2003}.

\begin{figure}
\begin{center}
\includegraphics[scale=0.49]{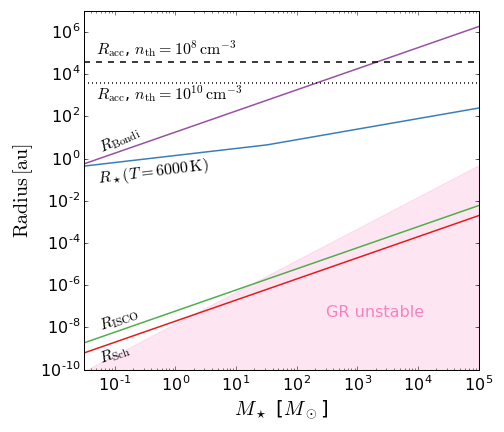}
\caption{
Characteristic scales related to the evolution of an accreting protostar:
stellar radius for a typical temperature of $T\simeq6000$ K (blue), Schwarzschild radius (red), radius of the innermost stable circular orbit (ISCO, green), and Bondi radius (purple). In addition we have included the accretion radius as defined in Equation (\ref{eq:racc}) for a density threshold of $\nth = 10^8\,\cmmm$ (black dashed) and $\nth = 10^{10}\,\cmmm$ (black dotted). The pink shaded area indicates the region of the parameter space where the star becomes GR unstable for $n=3$ polytropic stars \citep{Fricke_1973}.}
\label{fig:radius}
\end{center}
\end{figure}

Right after a protostar forms, the accretion timescale is shorter than the KH timescale.
In this accretion phase ($\tacc \lesssim \tkh$),  
we modify the left-hand-side of Equation (\ref{eq:energy}) and 
consider the accretion luminosity released at the stellar surface:
\begin{equation}
{G \mstar \mstardot \over \rstar}\simeq  4 \pi R_{\rm ph}^2 \sigma_{\rm SB} \tstar^4,
\label{eq:accretion_luminosity}
\end{equation}
where $R_{\rm ph}$ is the photospheric radius. 
In the early stages of the accretion phase, we assume $R_{\rm ph}\simeq 1.4~\rstar$ \citep{Stahler_1986}, which is derived for a spherically symmetric, quasi-steady model of an accreting protostar. Specifically, a freely-falling envelope is depositing material onto a growing hydrostatic core in an accretion shock. The latter is surrounded by a radiative precursor, transitioning into the optically thin envelope at the photosphere. Here, the factor of 1.4 is determined by $\HM$ opacity, but during later evolutionary stages,  other opacity effects like electron scattering will play a role. In accretion problems, other radii like the trapping radius, defined as the point where the radiative diffusion and dynamical timescales are of the same order, might be important. During the evolution of the system, the effective opacity can be written as a function of the opacity due to absorption ($\tau_{\rm abs}$) and to electron scattering ($\tau_{\rm es}$) as $\tau_{\rm eff} = \sqrt{\tau_{\rm abs}(\tau_{\rm abs} + \tau_{\rm es})}$. Inside the photosphere the gas is ionized and hence electron scattering dominates over absorption ($\tau_{\rm es} \gg \tau_{\rm abs}$). We can then estimate the trapping radius due to electron scattering as $R_{\rm tr}^{\rm es} = \mstardot \sigma_{\rm T} / 4\pi \mh c \simeq 4.45\,{\rm AU} \left( \mstardot \over \msun / {\rm yr} \right)$ \citep{Begelman_1978}. For our assumed accretion rate of $\mstardot \simeq 1\,\msunyr$, the trapping radius due to electron scattering is of the order of a few AU, which is consistent with the approximation $R_{\rm tr}^{\rm es} \lesssim R_{\rm ph} \sim \rstar$. On the other hand, we can also estimate the trapping radius due to $\HM$ absorption, given by equating the diffusion timescale and the free-fall time as $\tau_{\HM} R /c \simeq \sqrt{3\pi/32G\rho}$. To evaluate this expression we take the results from the one-zone model and solve for the radius. This gives us a value of $R_{\rm tr}^{\HM} \simeq 4\times 10^{-2}\,{\rm AU}$, which lies inside the photospheric radius at all times. As a result, we can adequately assume that the trapping radius might not influence the evolution of the object throughout the protostellar assembly. Future, self-consistent radiation-hydrodynamical simulations will provide a more complete understanding.
With these assumptions, we then evaluate the stellar radius
as a function of stellar mass $\mstar$, accretion rate $\mstardot$, and surface temperature $\tstar$:
\begin{align}
\rstar
\simeq &\,1.1
\left( \mstar \over \msun \right)^{1/3}
\left( \mstardot \over \msunyr \right)^{1/3} \left( \tstar \over 6000\,{\rm K}\right)^{-4/3} \,{\rm AU}.
\label{eq:rstar1}
\end{align}

Later in the evolution of the object, the internal luminosity from the star begins to dominate 
because the opacity decreases as the temperature increases inside the star ($\kappa \propto T^{-7/2}$),
resulting in $t_{\rm acc}>t_{\rm KH}$.
In a typical case for Population III star formation with a moderate accretion rate of $\mstardot \simeq 10^{-3}~\msunyr$,
the star contracts losing the thermal energy via radiative diffusion and forms a main-sequence star.
However, when the accretion rate is sufficiently high, the total luminosity (i.e., the accretion luminosity and the internal luminosity)
tends to exceed the Eddington luminosity during the KH contraction.
Then, the stellar surface expands in order to regulate the increase of the total luminosity, 
and the protostar evolves into a red-giant-like structure with a contracting core and an expanding envelope.
The critical accretion rate to bifurcate the protostellar evolution is estimated as $\mstardot \ga 4\times 10^{-3}~\msunyr$
\citep{Omukai_2003}.
According to stellar evolution calculations, even if $t_{\rm acc}>t_{\rm KH}$, 
the stellar surface continues to expand without contraction phases
when the accretion rate is higher than $\mstardot \ga 10^{-2}~\msunyr$ \citep{Hosokawa_2012}.
In this case, the stellar luminosity approaches the Eddington value for the corresponding mass.
Hence, the energy equilibrium equation can be written as
\begin{equation}
L_{\rm Edd,\star}= \frac{4 \pi G \mstar \mh c}{\sigma_{\rm T}}\simeq 4 \pi \rstar^2 \sigma_{\rm SB} \tstar^4 ,
\label{eq:internal_luminosity}
\end{equation}
from where we can derive an expression for the stellar radius 
during the expansion phase, as a function of the mass of the star and 
cthe surface temperature \citep{Hosokawa_2012, Hosokawa_2013}:
\begin{equation}
\rstar 
\simeq 0.78 \left(\mstar \over \msun \right)^{1/2} \left( \tstar \over 6000\,{\rm K}\right)^{-2} \,{\rm AU}.
\label{eq:rstar2}
\end{equation}

\begin{figure}
\begin{center}
\includegraphics[scale=0.5]{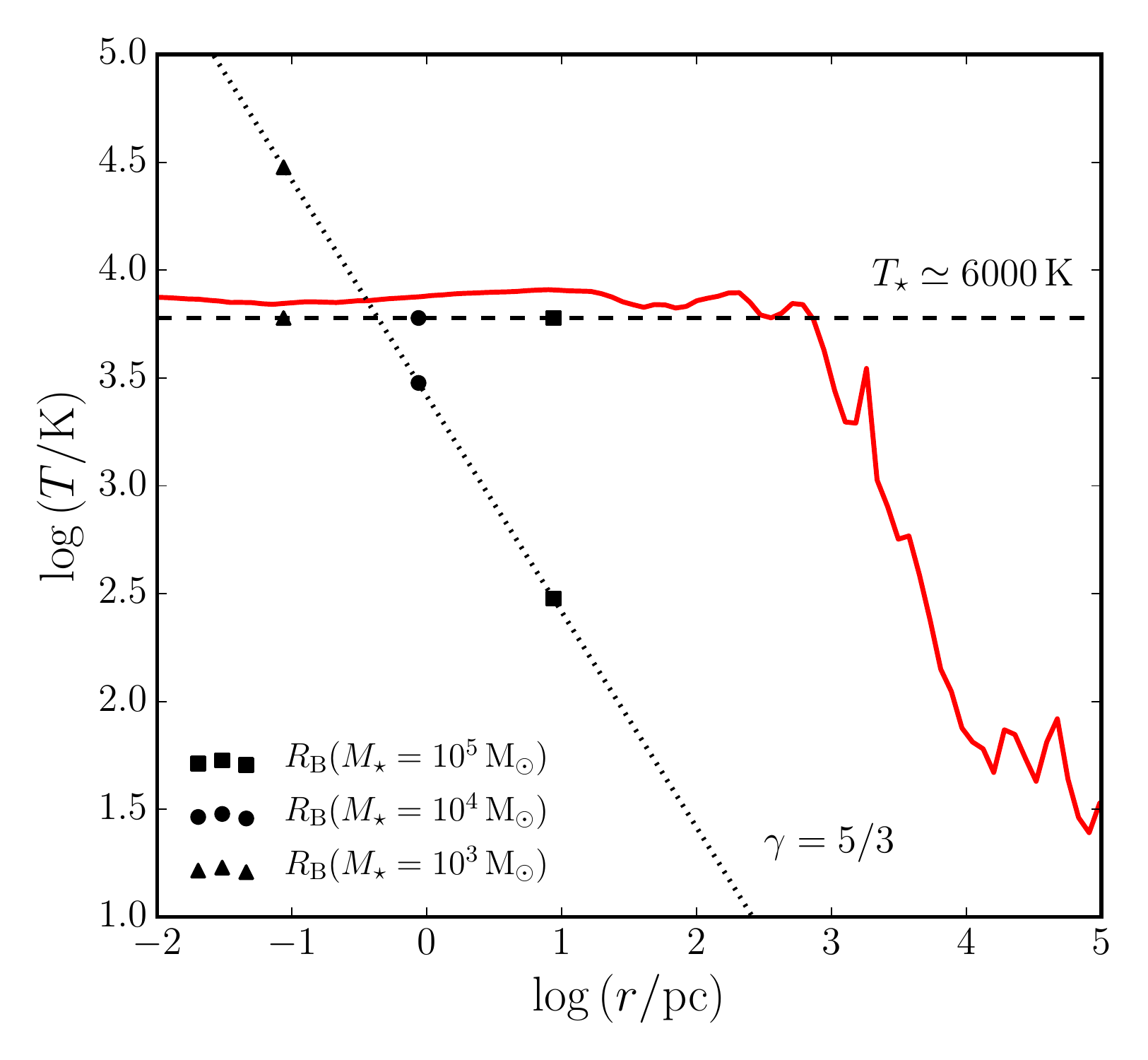}
\caption{Radial profiles of the temperature for our assumption $\tstar \simeq 6000\,{\rm K}$ (black dashed), our simulations (red solid, see Section \ref{subsec:cosmBCs}), and the adiabatic evolution of an optically thick object of mass $\mstar \simeq 10^4\,\msun$ (black dotted line). The assumption of a constant temperature is a good approximation up to scales $\simeq 100\,{\rm pc}$, which encloses the characteristic scales of the problem (see Figure \ref{fig:radius}). For reference, we have also included the value of the Bondi radius for masses $\mstar = 10^3$ (triangles), $10^4$ (circles), and $10^5\,\msun$ (squares).}
\label{fig:temp_profile}
\end{center}
\end{figure}

\begin{figure}
\begin{center}
\includegraphics[scale=0.48]{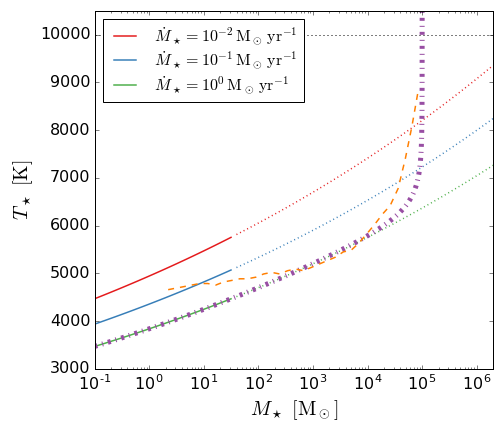}
\caption{Photospheric temperature as a function of stellar mass and accretion rate. Colors show the temperature evolution for $\mstardot = 10^{-2}$ (red solid), $10^{-1}$ (blue solid), $1\,\msunyr$ (green solid) based on Equation (23b) from \citet{Stahler_1986}. Solid lines represent the stage when $\tacc \lesssim \tkh$, while dotted lines are a rough extrapolation for higher masses. Additionally, we plot a time-dependent accretion rate of the form $\mstardot(t) = 1\,\msunyr e^{-t/t_{\rm ff,0}}$ (purple dash-dotted), as an illustrative case, and the results from \citet{Hosokawa_2013} (orange dashed). Note that, although both curves seem to agree well, the physical reasons for the rise in temperature are different (see text for more details). Black dotted line represents the ionizing temperature, $\Tion \simeq 10000\,{\rm K}$, at which the photosphere starts to emit non-negligible amounts of H ionizing radiation. The surface temperature of the protostar does not become high enough to start emitting hard UV radiation until quite late in the mass build-up. Hence, we can safely neglect its effect on the evolution of the central object early on.}
\label{fig:tstar}
\end{center}
\end{figure}

In Figure \ref{fig:radius} we compare the stellar radius (blue solid) for $\tstar \simeq 6000\,{\rm K}$ to other important length scales in our problem. Among them, we consider the Schwarzschild radius, $R_{\rm Sch} = 2G\mstar/ c^2$ (red solid), the innermost stable circular orbit (ISCO) radius, $R_{\rm ISCO} = 6G\mstar / c^2$ (green solid), and the Bondi radius, $R_{\rm B} = G\mstar / \css$ (purple solid). The protostar's radius grows from $\rstar \simeq 0.65 \,{\rm AU}$ at $\mstar \simeq 0.1 \msun$ to $\rstar \simeq 250 \,{\rm AU}$ at $\mstar \simeq 10^5 \msun$, well below the Bondi radius, but above both the Schwarzschild and ISCO radii for the whole range of masses. Additionally, the massive protostar does not enter the region where it becomes GR unstable (pink shaded region) during its evolution. This might indicate that the whole star does not collapse due to GR instability, but only its core, in agreement with \citet{Hosokawa_2013}. 

For the Bondi radius, we here assume for simplicity a nearly-constant sound speed, corresponding to $T(r)\simeq \tstar \simeq 6000\,{\rm K}$ (black dashed line in Figure \ref{fig:temp_profile}). This argument is based on the Larson-Penston collapse \citep{Larson_1969, Penston_1969}, which approximately describes the evolution of atomic cooling halos. In such a case, the density follows a profile $\rho \propto r^{-2}$ and hence the evolution of the temperature is nearly isothermal, up to radii of a few pc. Furthermore, we have verified that our simulations of the initial stages of collapse (see Section \ref{subsec:cosmBCs}) exhibit such near-isothermality out to $\simeq 100$\,pc, as shown by the red solid line in Figure \ref{fig:temp_profile}. It is clear that we oversimplify the situation here. In reality, the infalling matter will heat up when transitioning to optically-thick conditions inside the Bondi radius, and eventually follow an adiabatic evolution. In such a case, $\gamma = 5/3$ and the temperature has a radial dependency $T(r) \propto r^{-1}$ \citep{Shapiro_1983}, as represented for a stellar mass $\mstar \simeq 10^4\,\msun$ by the black dotted line in the same figure. In general, the transonic radius for Bondi accretion is given by $R_{\rm s} = \left(\frac{5-3\gamma}{4}\right) \frac{GM}{c_{{\rm s,}\infty}^2}$ \citep{Shapiro_1983}. If the gas evolution is characterized by a different $\gamma$ value \citep[e.g. $\gamma = 1.1$ for classical Population~III star formation,][]{Omukai_1998}, the Bondi radius might change by a factor of a few. Additionally, the transition to the adiabatic stage depends on how the diffusion and free-fall timescales compare to each other. Since the values for the trapping and photospheric radii are similar, this implies that gas outside the photospheric radius is not affected by this increase in temperature, further validating our assumption.

\subsection{Onset of radiative feedback}
\label{subsec:rad_feedback}
One factor that might dramatically influence the evolution described in this section  is the radiation emitted from the central accreting protostar. Here, we have assumed a constant surface temperature because of strong temperature dependence of H$^-$ opacity.
Throughout the evolution of the central object, however, its temperature will vary and eventually reach a point where radiative feedback becomes important. Previous studies have analyzed the formation of primordial supermassive stars in the rapid mass accretion regime and have found that the effective temperature of the object remains well below $10^4\,{\rm K}$, for protostellar masses up to $10^4\msun$, or so \citep{Hosokawa_2012, Hosokawa_2013}, suggesting that radiative feedback might not become important up to those mass scales.

It is useful to explore the likely temperature evolution of the growing supermassive protostar, in response to a realistic mass accretion history provided by a cosmological simulation. To this extent, we consider the photospheric temperature, given by the general stellar evolution calculations of \citet{Stahler_1986}:
\begin{equation}
T_\star \simeq 4000\,{\rm K} \left(\frac{{M}_\star}{{\rm M}_\odot}\right)^{0.044} \left(\frac{\dot{{M}}_\star}{{\rm M}_\odot\,{\rm yr}^{-1}} \right)^{-0.055}.
\label{eq:tstar}
\end{equation}
We have plotted this relation in Figure \ref{fig:tstar} for accretion rates of $10^{-2}$ (red), $10^{-1}$ (blue), and $1\,\msunyr$ (green). Because this relation is only valid when $\tacc \lesssim \tkh$, we have used solid lines up to the mass where this inequality inverts. For higher masses we make a rough extrapolation based on the same expression (dotted lines), although the evolution of the temperature for this stage is unclear. As it can be seen, the temperature at which the photosphere begins to emit H-ionizing radiation, $\Tion \simeq 10000\,{\rm K}$ (black dotted), is not reached in the range of accretion rates explored here. For lower values of $\mstardot$ the stellar radius is smaller than the photospheric one, and follows the zero age main sequence evolution. In such a case, the ionizing temperature can be reached well before $\mstar \simeq 10^4\,\msun$.

In addition to our constant accretion rate assumption, we have included a time-dependent toy model of the form $\mstardot(t) = 1\,\msunyr e^{-t/t_{\rm ff,0}}$ (purple dash-dotted line), with $t_{\rm ff,0} = 10^5\,{\rm yr}$ being the free-fall time in the core of the atomic cooling halo before the collapse \citep[e.g.][]{Safranek-Shrader_2012} and $\mstar(t=0) = \mf \simeq 0.1\,\msun$. We also compare with the \citet{Hosokawa_2013} results for the effective temperature $T_{\rm eff}$ at $\mstardot \simeq 1\,\msunyr$ (orange dashed line). Both models agree well, but in the former case the increase in the effective temperature is due to a drop in the accretion rate, while in the latter case this is a result of the decrease in the opacity because of the expansion of the stellar radius. Similar to the case of constant $\mstardot$, none of these models reach $\Tion$ during the evolution of the protostar up to masses of $10^5\,\msun$. Since the photospheric temperature at the characteristic accretion rate $\mstardot \simeq 1\,\msunyr$ never surpasses $10^4$\,K, we can safely neglect photo-ionization feedback from the central protostar during most of its evolution. This radiation only becomes important for accretion rates a few orders of magnitude lower than $1\,\msunyr$, or for masses $\mstar \simeq 10^5\,\msun$ in the case of our time-dependent model.

Lastly, the final mass of the protostar can be affected by continuum radiation driven mass loss once its total luminosity exceeds the Eddington luminosity \citep[e.g.,][]{Fiacconi_2016}, or by mass loss due to pulsations \citep{Inayoshi_2013}. As previously discussed, in the early stages of the evolution the total luminosity remains below the Eddington value and hence it is not affected by radiation-driven mass loss. However, as the mass of the protostar grows its luminosity increases and this scenario changes. In such case, the final mass of the protostar can vary significantly, as studied by \citet{Fiacconi_2016}. Furthermore, the accreting supermassive protostar might become pulsationally unstable, but the estimated mass loss rates are too low to effectively prevent protostellar growth \citep{Inayoshi_2013}. In summary, mass loss, either due to continuum radiation or pulsations, should not affect the early evolutionary stages, but continuum opacity might become important later on, when the protostellar mass approaches $\simeq 10^5 - 10^6\,\msun$ (see fifth panel of Figure \ref{fig:sketch}).

\begin{figure}
\begin{center}
\includegraphics[scale=1.25]{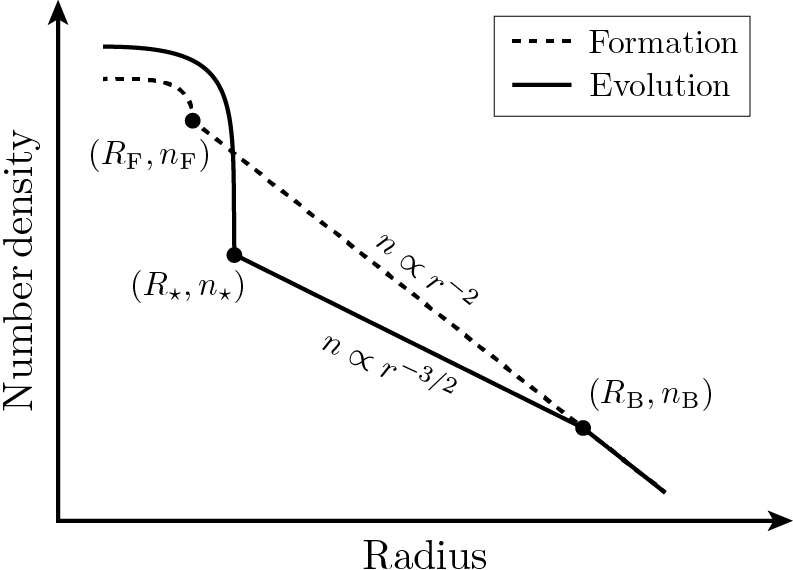}
\caption{Density-radius diagram for the protostar at the moment of formation (dashed) and its subsequent evolution (solid). When the fragment forms, its radius ($\rf$) is related to the Bondi radius ($\rB$) by an isothermal profile of the form $n \propto r^{-2}$. Once protostellar evolution starts, $\rB$ increases with mass (and hence with time) but the isothermal profile is kept outside it, while inside the relation changes to $n \propto r^{-3/2}$ down to the stellar radius $\rstar$.}
\label{fig:stellar_evolution}
\end{center}
\end{figure}

\begin{figure*}
\begin{center}
\includegraphics[scale=.43]{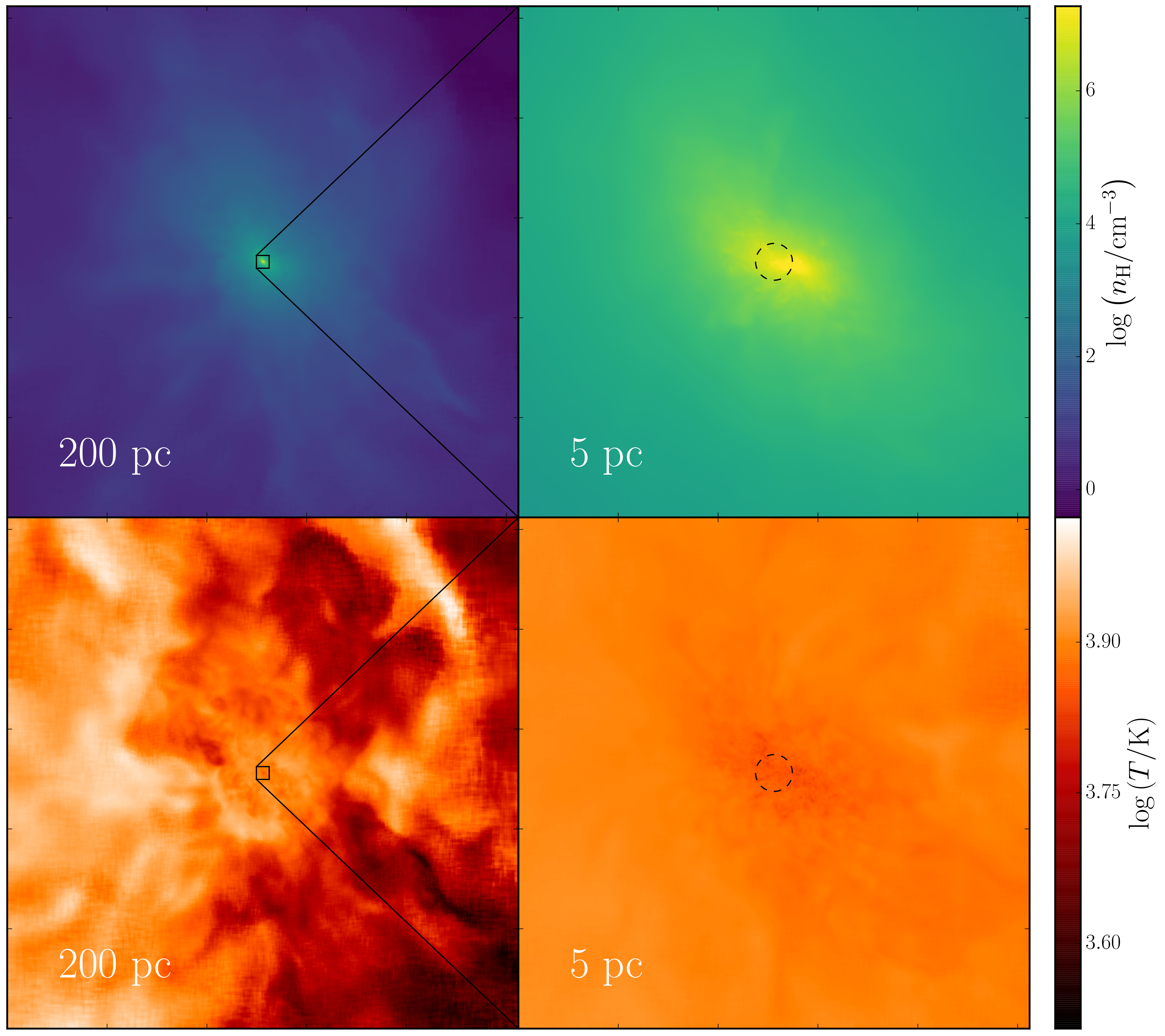}
\caption{Density ({\it top}) and temperature ({\it bottom}) projections of the central 200 (left) and 10 pc (right) for a low-resolution simulation of an atomic-cooling halo when the highest density cell first reaches $10^8\,\cmmm$. From Equation (\ref{eq:racc}), the accretion radius at this point is $\Racc \simeq 0.2\,{\rm pc}$, which is plotted in dashed black lines in both panels of the right column. At scales of 200 pc the cloud has an irregular morphology but it becomes nearly spherical on the smallest scales. The presence of turbulence can be deduced from the filamentary structure in the large-scale temperature map.}
\label{fig:sims}
\end{center}
\end{figure*}

\section{Lessons for sink algorithm}
\label{sec:sink}

\subsection{Accretion radius}
\label{subsec:acc_radius}

In the case of numerical simulations, the evolution of the central protostar requires either the implementation of sink particles \citep[e.g.][]{Latif_2013c} or an artificially stiffened equation of state \citep[e.g.][]{Hirano_2016}. For the former, we can use the treatment in the previous section of the protostellar evolution to construct a physically motivated sub-grid model.

The formation of the central object in the optically thick regime is characterized by a central fragment of density $\nf$ and radius $\rf$, and a isothermal profile of the form $n \propto r^{-2}$ outside that scale, as represented by the dashed line in the density-radius diagram in Figure \ref{fig:stellar_evolution}. The Bondi radius of the object is given by
\begin{equation}
\rB \simeq 17 \left( \mstar \over \msun \right) \left( T \over 6000 {\rm K} \right)^{-1}\,{\rm AU}.
\label{eq:rb}
\end{equation}
As seen from Figure \ref{fig:stellar_evolution}, $\rB$ is larger than $\rf$, and hence the relation between both quantities follow the isothermal profile, from which we can derive:
\begin{align}
\nB &= \nf \left( \rf \over \rB \right)^{2} \nonumber \\
&\simeq 4.66 \times 10^{13} f^{-2} \left( T \over 6000 {\rm K} \right)^3 \left( \mstar \over \msun \right)^{-2} \,\cmmm \mbox{\ ,}
\label{eq:nb}
\end{align}
where $\nB = n(r=\rB)$ is defined as the density at the Bondi radius.

The solid line in Figure \ref{fig:stellar_evolution} corresponds to the subsequent evolution, which is characterized by the growth of the stellar radius following Equation (\ref{eq:rstar2}) for the stage where $\tacc \gtrsim \tkh$ and accretion rates $\mstardot \gtrsim 10^{-2}\,\msunyr$. Outside the Bondi radius the evolution is still described by the isothermal profile $n \propto r^{-2}$, but inside $\rB$ material falls within a free-fall time, and hence it is represented by the relation $n \propto r^{-3/2}$ in the density-radius space. This allows us to relate the stellar and the Bondi radius as
\begin{align}
\nstar &= \nB \left( \rB \over \rstar \right)^{3/2}\\
&\simeq 5.13 \times 10^{15} f^{-1} \left( T \over 6000 {\rm K} \right)^{9/2} \left( \mstar \over \msun \right)^{5/4} \,\cmmm \mbox{\ ,}
\label{eq:nstar}
\end{align}
where $\nstar = n(r=\rstar)$ is defined as the density at the stellar radius.

Furthermore, inside the Bondi radius we can estimate the accretion rate of infalling gas as $\mdot_{\rm B} = 4 \pi \nB \mh \cs \rB^2$. Using Equation (\ref{eq:nb}) for $\nB$ and Equation (\ref{eq:rb}) for $\rB$, we derive: 
\begin{equation}
\mdot_{\rm B} \simeq 0.78 f^{-2} \left( T \over 6000 {\rm K} \right)^{3/2} \msunyr \mbox{\ ,}
\label{eq:mdotb}
\end{equation}

The framework here depicted can be used to derive a prescription for the accretion radius of the sink particle, $\Racc$. In the ideal case of a simulation with high enough resolution, we would set the accretion radius to $\rf$ or $\rstar$. Unfortunately this case is not always achievable and hence we need to choose $\Racc \lesssim \rB$  if the simulations resolves densities $n \gtrsim \nB$, with an accretion rate estimated by Equation (\ref{eq:mdotb}). On the other hand, the scenario becomes more complex when the simulation is only able to resolve a certain threshold density $\nth < \nB$. In such case, the choice of accretion radius should enclose the Bondi radius and, as suggested in Figure \ref{fig:stellar_evolution}, follow the isothermal profile. Hence, the value of the accretion rate for a given threshold density can be expressed as $\Racc = \rf \left( \nf / \nth \right)^{1/2}$.  

In summary, we can write a formula for the accretion radius as based on the threshold density as
\begin{equation}
\Racc \simeq
\begin{cases}
\vspace{1em}
17 \left( \mstar \over \msun \right) \left( T \over 6000 {\rm K} \right)^{-1}\,{\rm AU}, \hspace{1.6em} \text{if}\,\nth \gtrsim \nB \\
\left( 1.49 \times 10^{17} \,\cmmm \over \nth \right)^{1/2} f^{-1/2} \left( T \over 6000 {\rm K} \right)^{1/2}\,{\rm AU},\\
\hspace{12em} \text{if}\,\nth < \nB
\end{cases}
\label{eq:racc}
\end{equation}
For threshold densities of $\nth = 10^8 \,\cmmm$ and $\nth = 10^{10} \,\cmmm$, the initial values of the accretion radius in the low-mass regime (where the Bondi radius is not resolved) are given by $\Racc \simeq 3.8\times10^4\,{\rm AU}$ and $\Racc \simeq 3.8\times10^3\,{\rm AU}$, respectively. These values are kept until $\nB = \nth$, which occurs when the mass of the protostar is $\simeq 10^3\,\msun$ for the former and $\simeq 10^2\,\msun$ for the latter. From then on the accretion radius is given by the Bondi radius and the final mass is determined by the Bondi accretion rate, which is independent of the initial resolution of the simulation. Hence, the final mass does not depend on the choice for the threshold density.

The two initial accretion radii for $\nth = 10^8 \,\cmmm$ and $\nth = 10^{10} \,\cmmm$ are shown in Figure \ref{fig:radius} as black dashed line and black dotted line respectively. In both cases the accretion radius is larger than the stellar radius (calculated at $T=6000\,{\rm K}$) in the mass range $0.1 - 10^5 \,\msun$ once the sink is formed. As the star evolves, the Bondi radius increases with the mass of the star, eventually reaching the point when it is resolved, in which case the accretion radius should transition to $\Racc = \rB$. This method ensures that the central star, modeled as a sink particle, will always be enclosed by the accretion radius during its evolution, from its formation until it becomes a supermassive star.

Previous works have used different strategies to implement sink particles. For example, \citet{Latif_2013c, Shlosman_2016} assumed $\Racc = G\mstar / (\cs^2 + v_\infty^2)$ and then estimated the accretion rate as $\mstardot = 4\pi \rho_\infty \Racc \sqrt{1.2544\cs^2 + v_\infty^2}$, while \citet{Regan_2017} used a fixed accretion radius of four cells in the maximum refinement level. We expect all of these recipes to give a similar accretion rate of $\simeq 1 \msunyr$, which corresponds to our estimation from Equation (\ref{eq:mdotb}).

\subsection{Cosmological boundary conditions}
\label{subsec:cosmBCs}

We perform a low-resolution simulation of the collapse of an atomic cooling halo following a similar approach to the one described in \citet{Becerra_2015}. We start from cosmological initial conditions at redshift $z=99$ and box size of 2 Mpc (comoving) in a $\Lambda$ cold dark matter ($\Lambda$CDM) cosmology. We then follow the evolution of the halo until the highest density cell reaches the threshold density $\nth = 10^8\,\cmmm$. For that we have used a primordial chemistry network that includes five species (H, $\HH$, $\HM$, $\HP$, and $\e$) and cooling processes such as $\HM$ cooling, $\HH$ line cooling, $\HH$ collision-induced emission, Ly-$\alpha$ cooling, and inverse Compton cooling. The refinement criteria ensures that the Jeans length is resolved by at least 64 cells at every stage of the evolution.

We show the number density (top) and temperature (bottom) projections of the central object at scales of 200 (left) and 5 pc (right) at that instant in time in Figure \ref{fig:sims}. At large scales the cloud shows an irregular morphology, but it becomes nearly spherical at scales of $\simeq 10\,{\rm pc}$. This implies that the object reaches spherical symmetry at scales larger than the accretion radius at that point ($\Racc \simeq 0.2\,{\rm pc}$, as calculated in Section \ref{subsec:acc_radius}), which is plotted in black dashed lines.

\begin{figure}
\begin{center}
\hspace*{-0.2cm}\includegraphics[scale=0.5]{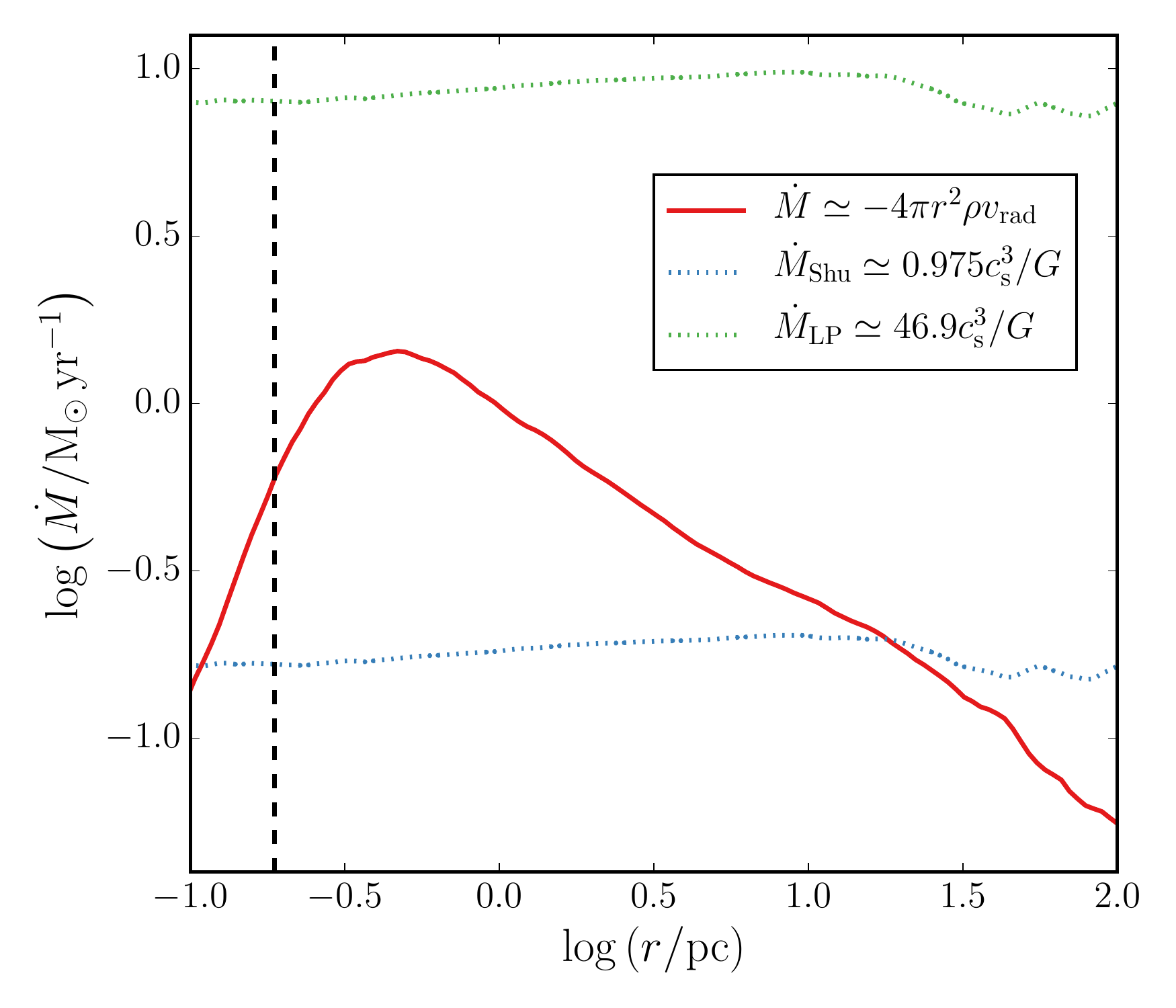}
\caption{Mass accretion rate as function of radius, centered on the highest density cell of the halo. The mass accretion rate reaches a maximum of $\mdot \simeq 1.4\,\msunyr$ at $r \simeq 0.5\,{\rm pc}$, and then it decreases to $\mdot \simeq 0.1\,\msunyr$ at a radial distance of $r \simeq 100\,{\rm pc}$. Note that the spatial non-constancy of $\mdot$ implies non-steady state conditions during the initial infall. At the accretion radius (vertical dashed line) the value of the mass infall rate is $\mdot \simeq 0.6\,\msunyr$. For comparison, we have also plotted $\mdot_{\rm Shu} \simeq 0.975\cs^3/G$ (blue dotted line), which stays between $0.1\,\msunyr$ and $0.2\,\msunyr$, and $\mdot_{\rm LP} \simeq 46.9\cs^3/G$ (green dotted line), which oscillates around $8.5\,\msunyr$.}
\label{fig:accretion}
\end{center}
\end{figure}

Finally, we analyze the accretion rate onto the object at the moment when the simulation reaches the threshold density in Figure \ref{fig:accretion}. The radial profile of the accretion rate is shown as red solid line, which is calculated as $\mdot = - 4 \pi r^2 \rho \vrad$, with $r$ the distance to the highest density cell, $\rho$ the mass density of hydrogen, and $\vrad$ the radial component of the velocity. For reference, we have also included the Shu accretion rate for spherical collapse, $\mdot_{\rm Shu} \simeq 0.975\cs^3/G$ \citep{Shu_1977}, and the Larson-Penston accretion rate for dynamical collapse, $\mdot_{\rm LP} \simeq 46.9\cs^3/G$ \citep{Larson_1969, Penston_1969}, as blue and green dotted lines, respectively. The mass accretion reaches a maximum of $\mdot \simeq 1.4\,\msunyr$ at $r \simeq 0.5\,{\rm pc}$ and then it decreases to values $\mdot \simeq 0.1\,\msunyr$ at larger scales, consistent with our estimation from Equation (\ref{eq:mdotb}). Up to scales of $\simeq$15 pc, its value lies in between the Shu and the Larson-Penston accretion rates, which remain roughly constant at $\mdot_{\rm Shu} \simeq 0.18\,\msunyr$ and $\mdot_{\rm LP} \simeq 8.5 \,\msunyr$ for the whole radial range. At the accretion radius (shown as a vertical black dashed line), the value of the mass infall rate is $\mdot \simeq 0.6\,\msunyr$, which is consistent with the values assumed throughout this study.

\subsection{Disk accretion}
\label{subsec:disk}

Throughout the paper, we discuss a sub-grid model within a sink assuming spherical symmetric accretion flows.
However, gas material with angular momentum form an accretion disk, through which the central protostar is fed.
In supermassive star formation, the disk would be unstable against its self-gravity
because of high accretion rates from the parent cloud ($\mstardot \sim 1~\msunyr$).
In such an unstable disk, the disk is likely to fragment into multiple clumps, which could migrate inward losing 
their orbital angular momentum due to gravitational interaction with the disk and other clumps.
Angular momentum redistribution induced by the clumps can drive the evolution of the disk and predict the formation of supermassive black holes in its nuclei as described by \citet{Lodato_2006}.
Eventually, most of the clumps can feed the gas into the central protostar episodically before the clumps evolve to main-sequence stars,
which could suppress the gas accretion through the disk due to ionizing radiation \citep{Inayoshi_Haiman_2014, Latif_2015}.
Moreover, the radius of the central protostar monotonically increases at $M_\star >10^2-10^3~\msun$ 
with an almost constant effective temperature of $T_{\rm eff}\simeq 5000~\K$, 
resulting in weak radiation feedback.
Since the average accretion rate through the disk is as high as $\sim 0.1~\msunyr$ and the duration of 
clump accretion episodes is shorter than the KH timescale at the stellar surface \citep{Sakurai_2016}, 
the evolution of the stellar structure is not affected by details of episodic accretion \citep{Sakurai_2015}.

\section{Summary and Conclusions}
\label{sec:conclusion}

In this paper we have developed a model for the early evolution of supermassive protostars. After the formation of the initial protostar the surrounding gas becomes optically thick to $\HM$ radiation, at which point we can robustly calculate the properties of the object using the equations of hydrostatic and thermal equilibrium. From that we obtain a characteristic density, radius, and mass of $\nf \simeq 4.6\times 10^{16}\,\cmmm$, $\rf \simeq 0.33\,{\rm AU}$, and $\mf \simeq 0.045\,\msun$, respectively, for a temperature $T=3000\,{\rm K}$.
An alternative approach to model the same situation is to use one-zone models. For that, we describe in detail the methods introduced in \citet{Inayoshi_2014} and provide explicit numerical fits for the $\HM$ cooling rate and opacity. Combined with the adiabatic heating rate, we can then calculate the critical density at which the gas becomes optically thick, which results in $n_{\rm crit} \simeq 2 \times 10^{16}\,{\rm cm}^{-3}$, consistent with the previous estimate. Hence we can robustly characterize the properties of the protostar in the initial optically thick regime.

The early stages of protostellar evolution, where $t_{\rm acc} \la t_{\rm KH}$, are described by the accretion of material onto the central object. For this case, we derive an expression for the protostellar radius as a function of the mass and accretion rate. Using a characteristic value of $\mstardot \simeq 1\,\msunyr$ for the accretion rate, we find that the protostellar radius grows as $\rstar \propto \mstar^{1/4}$ during this phase. Once internal sources of radiation start dominating, $\tkh \la \tacc$ and hence the radius-mass relation changes to $\rstar \propto \mstar^{1/2}$. For the case of a supermassive protostar, the radius varies from $\rstar \simeq 0.65 \,{\rm AU}$ at $\mstar \simeq 0.1 \msun$ to $\rstar \simeq 250 \,{\rm AU}$ at $\mstar \simeq 10^5 \msun$. For the surface temperature of the object, we base our analysis on the prescription of \citet{Stahler_1986}, deducing that it remains well below the ionizing temperature of $\Tion \simeq 10^4\,{\rm K}$ during most of its evolution. We can thus safely neglect UV ionizing radiation until the late stages of the assembly process.

In numerical simulations, supermassive protostars are commonly represented by sink particles. Our model allows us to derive the properties of such particles and implement a physically-motivated sub-grid model for their evolution in hydrodynamical codes. In particular, we derive an expression for the accretion radius ($\Racc$) as a function of the threshold density at which the sink particle is inserted ($\nth$), relating it to the physical conditions on the surface of the protostar. For high threshold densities our model proposes a numerical value based on the isothermal profile of the atomic cooling halo, but this value will eventually transition to $\Racc = \rB$ once the Bondi radius is resolved further in the evolution of the protostar. We can thus verify throughout the simulation that the accretion radius is well adjusted, in the sense that it is larger than the stellar radius at every moment during its evolution. 
Our new prescription for sink particles implies changes in the early stages of the evolution up to a protostar mass of $\simeq 10^{2-3}\,\msun$. After that, it follows the Bondi accretion scenario, and hence it has an accretion rate of $\simeq 1\,\msunyr$. The final mass of the object corresponds to $\simeq 10^{5-6}\,\msun$, similar to previous estimates in the literature.
We will track the accretion and KH timescales during the actual simulation to determine when the accretion rate becomes low enough and the star enters the KH phase. At that point, the radiation hydrodynamic effects from an ionizing central source would have to be taken into account.

The ultimate goal of this line of work is to simulate the assembly process of the first supermassive objects in the Universe in an ab-initio fashion. One key question then is: When will this build-up enter  a radiation-hydrodynamical phase, where the strong radiative feedback from the growing protostar will eventually turn the object into hyper-luminous beacons from the end of the cosmic dark ages? Those will be probed with next-generation observational facilities, such as the {\it James Webb Space Telescope (JWST)}, to be launched in 2018. An ideally complementary window into the formation of the first supermassive objects is provided by the gravitational wave signal accompanying the possible merger of binary black holes, which is a prime target for the planned Laser Interferometer Space Antenna (LISA). In light of this suite of next-generation facilities, simulations will have a key role  to play in providing physically robust predictions, based on well-motivated sub-grid prescriptions.

\section*{Acknowledgements}

We thank Kazuyuki Omukai for kindly providing permission to use the numerical models for $\HM$ cooling described in Section \ref{subsec:onezone}. 
KI acknowledges support by the Simons Foundation through the Simons Society of Fellows. VB was supported by NSF grant AST-1413501.
We also thank the anonymous referee for the constructive comments that helped to improve our paper.

\appendix

Following our discussion in Section \ref{subsec:onezone}, we can derive numerical fits to the $\HM$ cooling rate based on the description of Equations (\ref{eq:hmthin1}) to (\ref{eq:cooling_3}) as introduced by \citet{Inayoshi_2014}. The terms in the right-hand side of Equation (\ref{eq:hmthin2}) can be approximated as
$\Lhf = \khf n_{\rm HI} n_{\rm e}$ and $\Llf = \klf n_{\rm HI} n_{\rm e}$, respectively. The cooling rate coefficients are given by
\begin{align}
\klf &= 0.2345 \times \frac{T_3^{2.265}}{1+0.0360\,T_3^{2.149}}\times 10^{-28} \,{\rm erg}\,{\rm cm}^3\,{\rm s}^{-1}\\
\khf &= 10^{-27}\,T_3 \nonumber \\ 
    &\times \left( 1.4924 + 0.07815\,T_3 + 0.0063\,T_3^2 \right) \nonumber \\
    &\times  \left( 1 - 0.1535\,T_3^{0.5} \right)  \,{\rm erg}\,{\rm cm}^3\,{\rm s}^{-1},
\label{eq:kappafirst}
\end{align}
where $T_3 = T/10^3\,{\rm K}.$

Additionally, we estimate the Planck mean opacity for the $\HM$ free-free emission in both the low ($\klfffP$) and high ($\khfffP$) frequency regime, and for the $\HM$ bound-free emission for the high ($\khfbfP$) frequency regime as
\begin{align}
\klfffP &= \frac{10^{-28}}{T^3} \times 28.8 \times \frac{T_3^{-0.88}}{1 + 27.86\,T_3^{-2.15}} \,{\rm cm}^{-1} \\
\khfffP &= \frac{10^{-29}}{T^{2.5}} \exp{\left(-8.761 \over T_3 \right)} \times 2.868\,T_3^{-0.3326} \nonumber \\
&\times \left(1 + 2.544\,T_3^{1.1413} - 2.3369\,T_3^{1.162}\right) \,{\rm cm}^{-1} \\
\khfbfP &= \frac{10^{-11}}{T^{1.5}} \exp{\left(-8.761 \over T_3 \right)} \times 5.850 \nonumber \\
  &\times \left(1 - 0.1042\,T_3^{0.9419} + 0.0727\,T_3^{1.0278}\right) \,{\rm cm}^{-1}.
\end{align}

Another opacity source is the Rosseland mean opacity for $\HM$ free-free emission in the low frequency regime, which can be modeled as
\begin{align}
\klfffR &= 10^{-44}\,T^2 \times 0.4054\,T_3^{-8.180} \nonumber \\
  &\times \left(\frac{1 + 5.552\,T_3^{5.381} + 1.234\,T_3^{8.753}}{1 + 0.0296\,T_3^{1.180}}\right) \,{\rm cm}^{-1}.
\end{align}

Finally, we also estimate the opacity due to Rayleigh scattering for the high frequency regime as
\begin{align}
\khfRay &= 10^{-42} \exp{\left({8.761 \over T_3}\right)} \times T^{3.5} \times 1.206 \nonumber\\
  &\times \left(\frac{T_3^{0.750}}{1 + 0.139\,T_3^{1.204}}\right) \,{\rm cm}^{-1}
\label{eq:kappalast}
\end{align}

The total opacities for both regimes can then be written as $\khfR = \khfRay$, $\klfR = \klfffR$, $\khfP = \khfffP + \khfbfP$, and $\klfP = \klfffP$. With these approximations and Equation (\ref{eq:cooling_3}) we can rewrite the total $\HM$ cooling rate as a function of the optically thin cooling rates and the opacities:
\begin{equation}
\Lambda_{\rm H^-} = \frac{\Lhf}{1 + 3 \khfR \khfP \lambda_{\rm J} \lambda_{\rm J}}
+ \frac{\Llf}{1 + 3 \klfR \klfP \lambda_{\rm J} \lambda_{\rm J}}\mbox{\ .}
\label{eq:cooling}
\end{equation}
Here the characteristic length $\ell$ has been set to the Jeans length $\ell = \lambda_{\rm J}\simeq \cs \tff$, with $\cs = \sqrt{\gamma \kb T / \mu \mh}$ being the sound speed.

\end{document}